\documentstyle{ioplppt}
\begin{document}
\def\T{{\cal T}}

\jl{1}

\title{Hierarchy of Higher Dimensional Integrable System}

\author{YU Song-Ju\ftnote{1}{fpc30017@bkc.ritsumei.ac.jp}, 
Kouichi TODA\ftnote{2}{sph20063@bkc.ritsumei.ac.jp} and 
Takeshi FUKUYAMA\ftnote{4}{fukuyama@bkc.ritsumei.ac.jp}}

\address{Department of Physics, Ritsumeikan University, Kusatsu, Shiga 525 
JAPAN}

\begin{abstract}
Integrable equations in ($1 + 1$) dimensions have their own higher order 
integrable equations, like the KdV, mKdV and NLS hierarchies etc.  In this 
paper we consider whether integrable equations in ($2 + 1$) dimensions 
have also the analogous hierarchies to those in ($1 + 1$) dimensions.  
Explicitly is discussed the Bogoyavlenskii-Schiff(BS) equation.  For the 
BS hierarchy, there appears an ambiguity in the Painlev\'e test.  
Nevertheless, it may be concluded that the BS hierarchy is integrable.
\end{abstract}

\maketitle

\section{Introduction}
 
The inverse scattering transform(IST) method is a powerful tool for the 
investigation of ($1 + 1$) dimensional nonlinear differential equations.  
The extent of applicability of this method, however, depends on the 
dimension of space.  The well known ($2 + 1$) dimensional integrable 
equations are the Kadomtsev-Pettviashivili(KP), Davey-Stewarson(DS) 
equation and some new integrable equations which has been constructed by 
B.G.Konopelchenko and V.G.Dubrovsky\cite{kd}.  In this paper we consider 
a new integrable equation in (2+1) dimensions, the Bogoyavlenskii-Schiff 
equation (BS equation) \cite{b1,jschiff}.  This equation is given by

\begin{equation}
u_t + \Phi(u) u_z = 0,
\label{bs}
\end{equation}
where 

\begin{equation}
\Phi(u) \equiv \frac{1}{4} \partial^2_x + u + \frac{1}{2} u_x 
\partial^{-1}_x
\label{recursion}
\end{equation}
 is the recursion operator of KdV equation\cite{ff}.  
Potential form of this equation takes the form, 

\begin{equation}
\phi_{xt} + \frac{1}{4} \phi_{xxxz} + \phi_x \phi_{xz} + 
\frac{1}{2} \phi_{xx} \phi_z = 0 \hspace{0.5cm} (u \equiv \phi_x). 
\label{pbs}
\end{equation}
The BS equation has an infinite number of conservation laws\cite{b3}, the 
Painlev\'e property\cite{jschiff,b4} and  $N$ soliton 
solutions\cite{ytsf}.  That is, the BS equation is the ($2 + 1$) 
dimensional integrable equation.  For ($1 + 1$) dimensional case($z = x$), 
i.e. KdV equation, we can  construct the higher order equations of the 
KdV(the KdV hierarchy) with Lax operators and the conserved 
quantities\cite{l,zf}.  In this paper, we consider whether this 
construction of hierarchy is also valid for ($2 + 1$) dimensional 
integrable equation or not.  Namely the purpose of the present paper is to 
construct the BS hierarchy and to check its integrability through the 
Painlev\'e test.

In Sec.2, we review the hierarchy of an integrable equation in ($1 + 1$) 
dimensions and derive the Lax pair of the BS hierarchy.  In Sec.3, we 
investigate the integrability of the BS hierarchy obtained in the previous 
section.  Sec.4 is devoted to discussion mainly concerning with integrable 
equations in ($3 + 1$) dimensions.

\section{The Lax Representation of The BS Hierarchy}

The potential KdV equation 

\begin{equation}
\phi_{xt} + \frac{1}{4} \phi_{4x} + \frac{3}{2} \phi_x \phi_{xx} = 0 ,
\label{pkdV}
\end{equation}
admits the Lax representation\cite{LNPm10}

\begin{equation}
[L,T] = 0.
\label{laxeq1}
\end{equation}
$L$, $T$ operators have the form

\begin{eqnarray}
& &L(x,t) = \partial_x^2 + \phi_x(x,t), \label{L} \\ 
& &T(x,t) = \biggl(L(x,t)^{\frac{3}{2}}\biggr)_+ + \partial_t = 
\partial_x L(x,t) + T'(x,t) + \partial_t,
\label{T}
\end{eqnarray}
where $(~~)_+$ means the part of $(~~)$ with non-negative powers of 
$\partial_x$ and 

\begin{eqnarray}
L(x,t)^{\frac{1}{2}} &=& \partial_x + \frac{1}{2} \phi_x \partial_x^{-1} 
- \frac{1}{4} \phi_{xx} \partial_x^{-2} + \frac{1}{8} (\phi_{3x} - 
\phi_x^2) \partial_x^{-3} 
\nonumber \\
& & + \frac{3}{8} \biggl(\phi_x \phi_{xx} - \frac{1}{2} \phi{4x}\biggr) 
\partial_x^{-4} + \cdots, 
\label{L1/2} \\
T'(x,t) &=& \frac{1}{2} \phi_x \partial_x - \frac{1}{4} \phi_{xx}.
\label{T'}
\end{eqnarray}
The second equality in equation (\ref{T}) is given for the later use of 
the BS equation and others.

An infinite set of equations,

\begin{equation}
[L,T_n] = 0 \hspace{0.5cm} (n = 1, 2, 3, \cdots),
\label{laxhieeq1}
\end{equation}
constitute the potential KdV hierarchy and the equation for $n$ is called 
the $(2 n + 1)$th order potential KdV equation.  The operator $T_n$ has 
the form 

\begin{equation}
T_n(x,t) = \biggl(L(x,t)^{\frac{2 n + 1}{2}}\biggr)_+ + \partial_t = 
\partial_x L(x,t)^n + T'_n(x,t) + \partial_t.
\label{Tn}
\end{equation}
For $n = 2$ and $n = 3$, equation(\ref{laxhieeq1}) represents the $5$th 
and the $7$th order potential KdV equations, respectively: 

\begin{equation}
\fl \phi_{xt} + \frac{1}{16} \phi_{6x} + \frac{5}{8} \biggl(\phi_x^3 + 
\frac{1}{2} \phi_{xx}^2 + \phi_x \phi_{xx}\biggr)_x = 0
\label{5kdv}
\end{equation}
and

\begin{equation}
\fl T'_2 = \frac{1}{2} \phi_z \partial_x^3 - \frac{1}{4} \phi_{xx} 
\partial_x^2 + \biggl(\frac{1}{8} \phi_{3x} + \frac{7}{8} \phi_x^2\biggr) 
\partial_x - \frac{1}{16} \phi_{4x} - \frac{1}{8} \phi_x \phi_{xx}
\label{T'2}
\end{equation}
for $n = 2$.

\begin{equation}
\fl \phi_{xt} + \frac{1}{64} \phi_{8x} + \frac{7}{64} \biggl(5 \phi_x^4 
+ 10 \phi_x (\phi_{xx}^2 + \phi_x \phi_{3x}) + 2 \phi_x \phi_{5x} + 
4 \phi_{xx} \phi_{4x} + 3 \phi_{3x}^2\biggr)_x = 0
\label{7kdv}
\end{equation}
and 

\begin{eqnarray}
\fl & & T'_3 = \frac{1}{2} \phi_x \partial_x^5 - \frac{1}{4} \phi_{xx} 
\partial_x^4 + \biggl(\frac{1}{8} \phi_{3x} + \frac{11}{8} 
\phi_x^2\biggr) \partial_x^3 - \biggl(\frac{1}{16} \phi_{4x} - 
\frac{9}{8} \phi_x \phi_{xx}\biggr) \partial_x^2 + \biggl(\frac{1}{32} 
\phi_{5x} \nonumber \\
\fl & &- \frac{11}{32} \phi_{xx}^2 + \frac{16}{31} \phi_x \phi_{3x} + 
\frac{19}{16} \phi_x^3\biggr) \partial_x - \frac{1}{64} \phi_{6x} - 
\frac{9}{32} \phi_x \phi_{4x} - \frac{7}{16} \phi_{xx} \phi_{3x} + 
\frac{9}{32} \phi_x^2\phi_{xx}
\label{T'3}
\end{eqnarray}
for $n = 3$.  Equations (\ref{5kdv}) and (\ref{7kdv}) are 
integrable\cite{matsuno}.  

The Bogoyavlenskii's method is to modify the form of $T$ 
operator(\ref{T}) in ($1 + 1$) dimensions to in ($2 + 1$) dimensions as 

\begin{equation}
T(x,z,t) = \partial_z L(x,z,t) + \partial_t + T'(x,z,t), 
\label{BogoT}
\end{equation}
where

\begin{equation}
T'(x,z,t) = \frac{1}{2} \phi_z \partial_x - \frac{1}{4} \phi_{xz}.
\label{BogoT''}
\end{equation}
Namely we replace $\partial_x L$ in equation(\ref{T}) by $\partial_z L$.  
Here $L$ has the same form as equation(\ref{L}), though $\phi$, in this 
case, is a function of $x$, $z$ and $t$.  Then, the Lax 
equation(\ref{laxeq1}) becomes the BS equation (\ref{pbs})\cite{b1}.  We 
can construct the ($2 n + 1$)th order BS hierarchy analogously to the 
potential KdV hierarchy:

\begin{equation}
[L,T_n] = 0,
\label{Bogolaxhie}
\end{equation}
where $T_n$ have the forms 

\begin{equation}
T_n(x,z,t) = \partial_z L(x,z,t)^n + T'_n(x,z,t) + \partial_t.
\label{tn}
\end{equation} 
We could construct the same equations for the Lax 
equation(\ref{Bogolaxhie}) with other forms of the operator 
$T_n(x,z,t) = \underbrace{L \cdots L}_{\mbox{$m - 1$ times}} 
(\partial_z L) \underbrace{L \cdots L}_{\mbox{$n - m$ times}} + 
T''_n(x,z,t) + \partial_t$.  Here the operators $T''_n(x,z,t)$ is 
different from $T'_n(x,z,t)$, though we have the same equation after 
all.  For $n = 2$ and $n = 3$, equations(\ref{Bogolaxhie}) represents the 
$5$th and the $7$th order BS equations respectively: 

\hspace{-0.55cm}For $n = 2$, equation(\ref{Bogolaxhie}) becomes

\begin{eqnarray}
\fl &\phi_{xt}& + \frac{1}{16} \phi_{xxxxxz} + \frac{1}{2} \phi_x 
\phi_{xxxz} + \frac{3}{4} \phi_{xx} \phi_{xxz} + \frac{1}{2} \phi_{xxx} 
\phi_{xz} + \frac{1}{8} \phi_{xxxx} \phi_z + \phi_x^2 \phi_{xz} 
\nonumber \\ 
\fl &+& \frac{3}{4} \phi_x \phi_{xx} \phi_z + \frac{1}{8} \phi_{xx} 
\partial_x^{-1} (\phi_x^2)_z = 0,
\label{Bogo5th}
\end{eqnarray}
and

\begin{eqnarray}
\fl T'_2 &=& \frac{1}{2} \phi_z \partial_x^3 - \frac{1}{4} \phi_{xz} 
\partial_x^2 + \biggl(\frac{1}{8} \phi_{xxz} + \frac{3}{4} \phi_x \phi_z 
+ \frac{1}{8} \partial_x^{-1} (\phi_x^2)_z\biggr) \partial_x - 
\frac{1}{16} \phi_{xxxz}  \nonumber \\
\fl & & - \frac{1}{2} \phi_x \phi_{xz} + \frac{3}{8} \phi_{xx} \phi_z.
\label{BogoT'2}
\end{eqnarray}

\hspace{-0.55cm}In this case, the above equation(\ref{Bogo5th}) is 
equivalent to that obtained by Bogoyavlenskii\cite{b1,b2}.  For $n = 3$, 
equation(\ref{Bogolaxhie}) takes the form

\begin{eqnarray}
\fl &\phi_{xt}& + \frac{1}{64} \phi_{xxxxxxxz} + \frac{3}{16} \phi_x 
\phi_{xxxxxz} + \frac{15}{32} \phi_{xx} \phi_{xxxxz} + \frac{5}{8} 
\phi_{xxx} \phi_{xxxz} 
\nonumber \\
\fl &+& \frac{15}{32} \phi_{xxxx} \phi_{xxz} + \frac{3}{16} \phi_{xxxxx} 
\phi_{xz} + \frac{1}{32} \phi_{xxxxxx} \phi_z + \frac{9}{4} \phi_x 
\phi_{xx} \phi_{xxz} 
\nonumber \\
\fl &+& \frac{9}{8} \phi_{xx}^2 \phi_{xz} + \frac{3}{4} \phi_x^2 
\phi_{xxxz} + \frac{3}{2} \phi_x \phi_{xxx} \phi_{xz} + \frac{5}{16} 
\phi_x \phi_{xxxx} \phi_z + \frac{5}{8} \phi_{xx} \phi_{xxx} \phi_z 
\nonumber \\
\fl &-& \frac{1}{32} \phi_{xx} \partial_x^{-1} (\phi_{xx}^2)_z + 
\frac{1}{16} \phi_{xxxx} \partial_x^{-1} (\phi_x^2)_z + \phi_x^3 
\phi_{xz} + \frac{15}{16} \phi_x^2 \phi_{xx} \phi_z 
\nonumber \\
\fl &+& \frac{1}{16} \phi_{xx} \partial_x^{-1} (\phi_x^3)_z + 
\frac{3}{16} \phi_x \phi_{xx} \partial_x^{-1} (\phi_x)_z = 0,
\label{Bogo7th}
\end{eqnarray}
and

\begin{eqnarray}
\fl T'_3 &=& \frac{1}{2} \phi_z \partial_x^5 - \frac{1}{4} \phi_{xz} 
\partial_x^4 + \biggl(\frac{1}{8} \phi_{xxz} + \frac{5}{4} \phi_x 
\phi_{z} + \frac{1}{8} \partial_x^{-1} (\phi_x^2)_z\biggr) \partial_x^3 
\nonumber \\
\fl & & - \biggl(\frac{1}{16} \phi_{xxxz} + \frac{3}{4} \phi_x \phi_{xz} 
- \frac{15}{8} \phi_{xx} \phi_z\biggr) \partial_x^2 + \biggl(\frac{1}{32} 
\phi_{xxxxz} - \frac{5}{16} \phi_{xx} \phi_{xz} 
\nonumber \\
\fl & & + \frac{3}{8} \phi_x \phi_{xxz} - \frac{1}{32} \partial_x^{-1} 
(\phi_{xx})_z + \frac{25}{16} \phi_{xxx} \phi_{z} + \frac{15}{16} 
\phi_x^2 \phi_z + \frac{1}{16} \partial_x^{-1} (\phi_x^3)_z 
\nonumber \\
\fl & & + \frac{3}{16} \phi_x \partial_x^{-1} (\phi_x^2)_z\biggr) 
\partial_x - \frac{1}{64} \phi_{xxxxxz} - \frac{3}{16} \phi_x 
\phi_{xxxz} - \frac{1}{16} \phi_{xx} \phi_{xxz} - \frac{3}{8} \phi_{xxx} 
\phi_{xz} 
\nonumber \\
\fl & & + \frac{15}{32} \phi_{xxxx} \phi_z - \frac{3}{4} \phi_x^2 
\phi_{xz} + \frac{15}{16} \phi_x \phi_{xx} \phi_z + \frac{3}{32} 
\phi_{xx} \partial_x^{-1} (\phi_x^2)_z.
\label{BogoT'3}
\end{eqnarray}
Equations(\ref{Bogo5th}) and (\ref{Bogo7th}) are, of course, reduced to 
the $5$th and $7$th order potential KdV equations in the case of $z = x$, 
respectively.  The calculations for $n \ge 4 $ can be performed similarly, 
but the procedure to determine $T'_n$ operators for larger $n$ becomes 
more complicated.  

\section{Application of The Painlev\'e Test}

We have constructed the BS hierarchy in the previous section.  In this 
section we proceed to check the integrability of the BS 
hierarchy(\ref{Bogolaxhie}) through the Painlev\'e test in the sense of 
WTC\cite{wtc,t}.  For the BS hierarchy, there appears an ambiguity in this 
test.  Hence we give the detail of calculation, which may help to 
understand what it is.  In order to eliminate the integral operator, we 
rewrite the $5$th order BS equation(\ref{Bogo5th}) in the form of coupled 
systems, 

\begin{eqnarray}
&\phi_{xt}& + \frac{1}{16} \phi_{xxxxxz} + \frac{1}{2} \phi_x 
\phi_{xxxz} + \frac{3}{4} \phi_{xx} \phi_{xxz} + \frac{1}{2} \phi_{xxx} 
\phi_{xz} 
\nonumber \\ 
& & + \frac{1}{8} \phi_{xxxx} \phi_z - \rho_{xx} \phi_{xz} + \frac{3}{4} 
\phi_x \phi_{xx} \phi_z - \frac{1}{8} \phi_{xx} \rho_{xz} = 0, 
\label{Bogo5th1} \\
&\rho_{xx}& + \phi_x^2 = 0.
\label{Bogo5th2}
\end{eqnarray}
We now consider a local Laurent expansion in the neighborhood of a 
non-characteristic singular manifold $\gamma(x,z,t) = 0$, 
($\gamma_x \ne 0$).  Let us assume that the leading orders of the 
solutions of equations(\ref{Bogo5th1}) and (\ref{Bogo5th2}) have the 
forms

\begin{equation}
\phi = \phi_0 \gamma^{\alpha}, \hspace{0.5cm} \rho = \rho_0 
\gamma^{\beta}.
\label{leadingsols}
\end{equation}
Here $\phi_0$ and $\rho_0$ are some analytic functions.  Substituting
(\ref{leadingsols}) into (\ref{Bogo5th1}) and (\ref{Bogo5th2}), and 
equating the powers of the most dominant terms, we obtain 

\begin{equation}
\alpha = -1, \hspace{0.5cm} \beta = -2
\label{leadingorder}
\end{equation}
with

\begin{equation}
\phi_0 (\phi_0 - 6 \gamma_x) (\phi_0 - 2 \gamma_x) = 0, \hspace{0.5cm} 
6 \rho_0 + \phi_0^2 = 0.
\label{leadingterm}
\end{equation}
To find the resonances we now substitute the full Laurent expansion of 
the solutions

\begin{equation}
\phi = \sum_{j = 0} \phi_j \gamma^{j - 1}, \hspace{0.5cm} \rho = 
\sum_{j = 0} \rho_j \gamma^{j - 2},
\label{Painlevesols}
\end{equation}
into equations(\ref{Bogo5th1}) and (\ref{Bogo5th2}).  Rearranging 
equation(\ref{Bogo5th1}) into terms of $\gamma^{j - 7}$ and the other 
higher powers of $\gamma$ ($f_j$), we obtain recurrence relations for 
$\phi_j$, $\rho_j$, 

\begin{eqnarray}
& & \biggl(\bigl[(j - 1) (j - 2) (j - 3) (j - 4) (j - 5) (j - 6) 
\gamma_x^5 \gamma_z 
\nonumber \\
& & \hspace{0.6cm} - 10 (j - 1) (j - 6) (j^2 - 7 j + 16) \gamma_x^4 
\gamma_z \phi_0 
\nonumber \\
& & \hspace{0.6cm} - 108 (j - 1) (j - 2) \gamma_x^3 \gamma_z \rho_0 + 
12 (j - 1) (j - 6) \gamma_x^3 \gamma_z \phi_0^2\bigr] \phi_j 
\nonumber \\
& & \hspace{0.6cm} - 36 (j - 2) (j - 3) \gamma_x^3 \gamma_z \phi_0 
\rho_j\biggr) \gamma^{j - 7} = f_j. 
\label{gammaj-71}
\end{eqnarray}
Similarly rearranging equation(\ref{Bogo5th2}) into terms of 
$\gamma^{j - 4}$ and higher powers of $\gamma$ ($g_j$), we have 

\begin{equation}
\bigl(- 2 (j - 1) \gamma_x^2 \phi_0 \phi_j + (j - 2) (j - 3) \gamma_x^2 
\rho_j\bigr) \gamma^{j - 4} = g_j.
 \label{gammaj-72}
\end{equation}
Here $f_j$ and $g_j$ are given in terms of $\phi_l$ and $\rho_l$ 
($0 \le l \le j - 1$).  Then we get three types of resonances: 

\begin{eqnarray}
& & j = 0, 1, 1, 2, 3, 4, 5, 6, 
\label{resonance1} \\
& & j = -1, 1, 2, 2, 3, 5, 6, 8
\label{resonance2}
\end{eqnarray}
and

\begin{equation}
j = -3, -1, 1, 2, 3, 6, 8, 10.
\label{resonance3}
\end{equation}
These correspond to the choices of solutions of equations
(\ref{leadingterm}),

\begin{eqnarray}
& & \phi_0 = 0, \hspace{0.5cm} \rho_0 = 0, 
\label{sol1} \\
& & \phi_0 = 2 \gamma_x, \hspace{0.5cm} \rho_0 = - (2/3) \gamma_x^2 
\label{sol2}
\end{eqnarray}
and

\begin{equation}
\phi_0 = 6 \gamma_x, \hspace{0.5cm} \rho_0 = - 6 \gamma_x^2,
\label{sol3}
\end{equation}
respectively.  Here the first choice (\ref{sol1}) does not include the 
universal resonance $j = - 1$ and is discarded.  If the recurrence 
relations are consistently satisfied at the resonances then the 
differential equations is said to possess the Painlev\'e property.  

Succeeding coefficients $\phi_j$ and $\rho_j$ are determined from 
equations (\ref{gammaj-71}) and (\ref{gammaj-72}).  However, from the 
consistency condition, they must include arbitrary functions at the 
resonances.  To simplify the calculations, we use the reduced manifold 
ansatz of Kruskal\cite{kjh}.  That is, $\gamma(x,z,t) = x + \delta(z,t)$ 
and $\phi_j$, $\rho_j$ are function of $z$ and $t$.  

Equations (\ref{gammaj-71}) and (\ref{gammaj-72}) must be satisfied in 
the respective powers of $\gamma$.  The lowest powers of $\gamma$, 
($\gamma^{-7}$,$\gamma^{-4}$) in equations((\ref{gammaj-71}),
(\ref{gammaj-72})) give 
 
\begin{equation}
\phi_0 (\phi_0 - 2) (\phi_0 - 6) = 0, \hspace{0.5cm} \rho_0 = - 
\frac{1}{6} \phi_0^2,
\label{gammam7m4}
\end{equation}
which constitutes a set of equations(\ref{sol1}), (\ref{sol2}), and 
(\ref{sol3}).  Higher powers of $\gamma$,($\gamma^{-7 + k}$,
$\gamma^{-4 + k}$ with positive integer $k$)  in 
equations((\ref{gammaj-71}),(\ref{gammaj-72})) lead us to the following 
consistency conditions.

\begin{equation}
\fl (\gamma^{-6}, \gamma^{-3}) :~~~~~~\phi_0 \rho_1 \delta_z = 0, 
\hspace{0.5cm} \rho_1 = 0.
\label{gammam6m3}
\end{equation}
Hence $\phi_1$ is arbitrary.  

\begin{equation}
\fl (\gamma^{-5}, \gamma^{-2}) :~~~~~~\phi_0 \bigg[(\phi_0 - 2) 
\phi_{1,z} + 2 (\phi_0 - 5) \phi_2 \delta_z\bigg] = 0, 
\hspace{0.5cm} \phi_0 \phi_2 = 0,
\label{gammam5m2}
\end{equation}
These equations imply that $\phi_0$ is fixed as $2$ and $\rho_2$ is 
arbitrary, which contradicts equation(\ref{sol3}).  So we must choose 
(\ref{resonance2}).  

\begin{equation}
\fl (\gamma^{-4}, \gamma^{-1}) :~~~~~~(3 \phi_0 - 20) \phi_3 \delta_z = 0, 
\hspace{0.5cm} \phi_3= 0, 
\label{gammam4m1}
\end{equation}
and $\rho_3$ is arbitrary.  

\begin{equation}
\fl (\gamma^{-3}, \gamma^0) :~~~~~~6 (13 \phi_4 - 3 \rho_4) \delta_z + 
8 \delta_t - \rho_{3,z} = 0, \hspace{0.5cm} \rho_4 = 6 \phi_4.
\label{gammam30}
\end{equation}

\begin{equation}
\fl (\gamma^{-2}, \gamma^1) :~~~~~~9 (8 \phi_5 - 3 \rho_5) \delta_z + 
(6 \phi_4 - \rho_4)_z = 0, \hspace{0.5cm} \rho_5 = \frac{8}{3} \phi_5
\label{gammam21}
\end{equation}
and 

\begin{equation}
\fl (\gamma^{-1}, \gamma^2) :~~~~~~36 (5 \phi_6 - 3 \rho_6) \delta_z + 
(8 \phi_5 - 3 \rho_5)_z = 0, \hspace{0.5cm} \rho_6 = \frac{5}{3} \phi_6.
\label{gammam12}
\end{equation}
It follows from equations (\ref{gammam21}) and (\ref{gammam12}) that one of 
the two variables must be arbitrary in both pairs ($\phi_5$, $\rho_5$) and 
($\phi_6$, $\rho_6$).  

\begin{equation}
\fl (\gamma^0, \gamma^3) :~~~~~~12 (13 \phi_7 - 15 \rho_7) \delta_z - 
(5 \phi_6 + 4 \rho_6)_z  - 6 \phi_5 \phi_{1,z} = 0, \hspace{0.5cm} 
\rho_7 = \frac{6}{5} \phi_7.
\label{gamma03}
\end{equation}

\begin{eqnarray}
\fl & & (\gamma^1, \gamma^4) :~~~~~~\bigg[(504 \phi_8 - 540 \rho_8) + 
72 \phi_4^2\bigg] \delta_z + 3 \phi_4 (- 18 \rho_4 \delta_z + 
8 \delta_t - \rho_{3,z}) \nonumber \\
\fl & & \hspace{2.5cm} + 2 (6 \phi_7 - 5 \rho_7)_z = 0, \nonumber \\
\fl & & \hspace{2.2cm} \rho_8 = \frac{14}{15} \phi_8 - 
\frac{3}{10} \phi_4^2.
\label{gamma14}
\end{eqnarray}
That is, one of the two variables ($\phi_8$, $\rho_8$) must be arbitrary.  
Problematic is the following point:  $\phi_2$ at double root $j = 2$ has 
been  fixed as $0$ and the system lacks one more arbitrary function.  
However, $\phi_j$ and $\rho_j$ are determined consistently and, thus, 
dependent variables $\phi$ and $\rho$ have only movable poles.  The 
apparently  similar situation occurred in the system of coupled non-liner 
Sch\"odinger equation(NLS)\cite{stl}.  Starting from NLS leaving the 
constant coefficients as free parameters in \cite{stl}, Sahadevan et al. 
these parameters so as to include $N$ arbitrary functions at $N$-tuply 
degenerate resonance.  However, there seems to exist an essential 
difference between their NLS and our $5$th order BS equation since NLS 
does not allow the consistent Laurent expansion except for the special 
choices of constant coefficients above mentioned.  In the $5$th order BS 
equation there is no such room to adjust the number of arbitrary function, 
though the $5$th order BS equation has no inconsistency.  We want to add 
one remark here.  The coupled system of equations (\ref{Bogo5th1}) and 
(\ref{Bogo5th2}) is reduced to the $5$th order potential KdV equation in 
the case of $z = x$.  Even in that case, the resonance structures (the 
positions and the number of arbitrary functions etc.) are quite the same 
as the $5$th order BS equation.  The $5$th order potential KdV equation is 
manifestly integrable.  From these facts we may conclude that the $5$th 
order BS equation is also integrable.

Analogously, we have checked that the $7$th order BS equation is integrable 
in the same sense.

Higher order equation can be constructed in several different ways.  All 
the higher order BS equations(\ref{Bogolaxhie}) are rewritten as 

\begin{equation}
u_t + \Phi(u)^n u_z = 0
\label{recBogolaxhie}
\end{equation}
by the use of the recursion operator of KdV equation(\ref{recursion}).  
Equation(\ref{recBogolaxhie}) was conjectured to be 
integrable\cite{jschiff} because the integrability arose from the 
existence of the recursion operator\cite{fs}.  However, the recursion 
operator in ($2 + 1$) dimensions is not so clear as in the case of 
($1 + 1$) dimensions.  Explicit calculation in this section has uncovered 
a new feature and has shown that the BS hierarchy is integrable.

\section{Discussion}

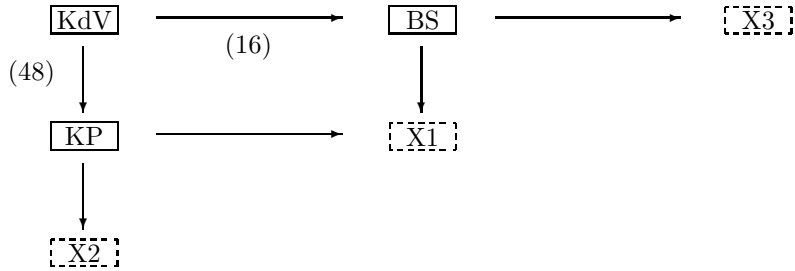
\begin{figure}[h]
\begin{center}
\begin{picture}(100,80)
\put(-100,70){\framebox(25,10){KdV}}
\put(-60,76){\vector(2,0){70}}
\put(28,70){\framebox(25,10){BS}}
\put(68,76){\vector(2,0){70}}
\put(155,70){\dashbox{2.0}(25,10){X3}}
\put(-120,50){\makebox(25,10){(\ref{KPL})}}
\put(-88,65){\vector(0,-1){25}}
\put(-38,60){\makebox(25,10){(\ref{BogoT})}}
\put(40,65){\vector(0,-1){25}}
\put(-100,26){\framebox(25,10){KP}}
\put(-60,32){\vector(2,0){70}}
\put(28,26){\dashbox{2.0}(25,10){X1}}
\put(-88,21){\vector(0,-1){25}}
\put(-100,-18){\dashbox{2.0}(25,10){X2}}
\end{picture}
\end{center}
\caption{Scheme of extensions of the KdV equation.  There are two 
directional routes of extensions: One leads us to the BS equation and the 
other does to the KP equation.  X equations are possible ($3 + 1$) 
dimensional integrable equations obtained from further extensions.}
\label{extensions}
\end{figure}

In this paper, we have constructed the BS hierarchy(\ref{Bogolaxhie}) 
using the Bogoyavlenskii's method.  This hierarchy is represented by the 
recursion operator(\ref{recursion}) and reduced to the potential KdV 
hierarchy by setting $z = x$.  The BS hierarchy(\ref{Bogolaxhie}), 
(\ref{recBogolaxhie}) has been proved to be integrable through the 
Painlev\'e test.  The BS equation is an extension of the KdV equation 
like the KP equation is.  However their extensions differ in algebraic 
structure, which is schematically depicted in Figure.\ref{extensions}.

\hspace{-0.41cm} The potential KP equation is obtained from the potential 
KdV equation by the replacement,

\begin{eqnarray}
L(x,t) \longrightarrow L'(x,y,t) &=& \partial_x^2 + \phi_x(x,y,t) + 
\partial_y \nonumber \\
&\equiv& L(x,y,t) + \partial_y,
\label{KPL}
\end{eqnarray}
The form of $T(x,y,t)$ is 

\begin{equation}
T(x,y,t) = \partial_x L(x,y,t) + T'(x,y,t) + \partial_t,
\label{KPT}
\end{equation}
where

\begin{equation}
T'(x,y,t) = \frac{1}{2} \phi_x \partial_x - \frac{1}{4} \phi_{xx} - 
\frac{3}{4} \phi_y.
\label{KPT'}
\end{equation}

Here we have denoted an extended spatial coordinate as y.  These 
extensions are contrasted with those of the BS equation(\ref{BogoT}) 
in which explicit $y$ (in that case $z$) dependence has appeared not in 
$L$ but in $T$.

These two extensions may be performed in the different spatial 
dimensions (i.e. $y \ne z$),  which is described by X1 in 
Figure.\ref{extensions}.  For instance, we consider the Lax 
equation(\ref{laxeq1}) with the operators $L'$ and $T$ having the form

\begin{eqnarray}
& &L'(x,y,z,t) = \partial_x^2 + \phi_x(x,y,z,t) + \partial_y 
\equiv L(x,y,z,t) + \partial_y, 
\label{newL} \\
& &T(x,y,z,t) = \partial_z L(x,y,z,t) + T'(x,y,z,t) + \partial_t,
\label{newT}
\end{eqnarray}
where

\begin{equation}
T'(x,y,z,t) = \frac{1}{2} \phi_z \partial_x - \frac{1}{4} \phi_{xz} 
- \frac{1}{4} \partial_x^{-1} \phi_{yz} + c \partial_z^2
\label{newT'}
\end{equation}
and $c$ is a constant.  However, the Lax equation(\ref{laxeq1}) with 
operators(\ref{newL}) and (\ref{newT}) in ($3 + 1$) dimensions is reduced 
to ($2 + 1$) dimensional equation, 

\begin{equation}
\phi_{xt} + \frac{1}{4} \phi_{xxxz} + \phi_x \phi_{xz} + \frac{1}{2} 
\phi_{xx} \phi_z + c^2 \partial_x^{-1} \phi_{zzz} = 0
\label{neweqfty}
\end{equation}
with the condition, 

\begin{equation}
z = 2 c y. 
\label{reducedcondition}
\end{equation}
This condition comes from the requirement that the coefficient of 
$\partial_z$ should be vanished in the Lax equation and, therefore, seems 
to be indispensable.  Equation(\ref{neweqfty}) in ($2 + 1$) dimensions is 
the same as that obtained by Bogoyavlenskii\cite{b1}.  We have checked 
that equation(\ref{neweqfty}) passes the Painlev\'e test.  By the 
dependent variable transformation $\phi = 2 (\log \tau)_x$, equation 
(\ref{neweqfty}) is transformed into the trilinear form,

\begin{equation}
(36 \T_x^2 \T_t + \T^4_x \T^{\ast}_z + 8 \T_x^3 \T^{\ast}_x \T_z 
+ 9 \T_z^3) \tau \cdot \tau \cdot \tau = 0.
\label{trineweqfty}
\end{equation}
The operators $\T$, $\T^{\ast}$ are defined by\cite{grh,hgr}

\begin{equation}
\fl \T_z^n f(z) \cdot g(z) \cdot h(z) \equiv (\partial_{z_1} + 
j \partial_{z_2} + j^2 
\partial_{z_3})^nf(z_1)g(z_2)h(z_3)|_{z_1=z_2=z_3=z}, \label{Tdef}
\end{equation}
where $j$ is the cubic root of unity, $j = \exp(2 \i \pi/3)$. 
$\T^{\ast}_z$ is the complex conjugate operator of $\T_z$ obtained 
by replacing $(\partial_{z_1} + j \partial_{z_2} + j^2 
\partial_{z_3})$ by $(\partial_{z_1} + j^2 \partial_{z_2} + j 
\partial_{z_3})$.  Equation(\ref{trineweqfty}) coincides with the 
trilinear form coming from the quite different approach\cite{hgr}.  
Namely, Hietarinta, Grammaticos and Ramani derived this equation from the 
singularity analysis of trilinear equations.  We have obtained $N$ soliton 
solutions of equation(\ref{trineweqfty}) by the computer program 
{\it Mathematica} up to $N = 5$.  However, we can not construct the 
hierarchy of equation(\ref{neweqfty}).  Namely we can not determine 
$T'(x,y,z,t)$ operator in 

\begin{equation}
T(x,y,z,t) = \partial_z L(x,y,z,t)^n + T'(x,y,z,t) + \partial_t,
\label{newThie}
\end{equation}
so as to constitute Lax equation(\ref{laxeq1}).  

A little bit different formulation to reach X1 candidate was proposed by 
us\cite{ytsf},

\begin{equation}
\biggl(u_t + \Phi(u) u_z\biggr)_x + \frac{3}{4} u_{yy} = 0. 
\label{newoldeq}
\end{equation}
Unfortunately, equation (\ref{newoldeq}) has a movable logarithmic 
branch point.  

One of the examples for ($3 + 1$) dimensional X3 equations in 
Figure.\ref{extensions} is obtained also by Bogoyavlenskii\cite{b4}.  
However, it is not affirmative unless it is proved to be integrable.  

The respective routes in Figure.\ref{extensions} are not unique.  
Different formulations at each path are possible, though we can not 
specify which is correct at the present stage.  Such specification 
enforces us to clarify the algebraic structures and integrability of the 
equations depicted in Figure.\ref{extensions} and their hierarchies 
furthermore.

\ack

We would like to thank Professor T. Kawahara for useful discussions.


\end{document}